\documentclass[12pt]{article}

\usepackage[totalheight = 23cm, totalwidth = 17cm]{geometry}
\usepackage{amssymb,amsmath,amsfonts,amsbsy,axodraw}

\def\mathbi#1{\textbf{\em #1}}

\newcommand{\mpl}{m_{\rm Pl}}
\newcommand{\calL}{{\cal L}}
\newcommand{\calO}{{\cal O}}
\newcommand{\calP}{{\cal P}}
\newcommand{\calR}{{\cal R}}

\begin{document}

\begin{titlepage}

\rightline{\footnotesize{APCTP-Pre2014-013}}

\begin{center}

\vskip 1.0cm

\textbf{\Huge Running of scalar spectral index \\ in multi-field inflation}

\vskip 1.0cm

\large{
Jinn-Ouk Gong$^{a,b}$
}

\vskip 0.5cm

\small{\it 
$^{a}$Asia Pacific Center for Theoretical Physics, Pohang 790-784, Korea 
\\
$^{b}$Department of Physics, Postech, Pohang 790-784, Korea
}

\vskip 1.2cm

\end{center}

\begin{abstract}

We compute the running of the scalar spectral index in general multi-field slow-roll inflation. By incorporating explicit momentum dependence at the moment of horizon crossing, we can find the running straightforwardly. At the same time, we can distinguish the contributions from the quasi de Sitter background and the super-horizon evolution of the field fluctuations.

\end{abstract}

\end{titlepage}

\setcounter{page}{0}
\newpage
\setcounter{page}{1}

\section{Introduction}

The recent developments in the observations on the cosmic microwave background (CMB) have constrained the main cosmological parameters for the simplest $\Lambda$CDM model with great accuracy~\cite{Ade:2013zuv}. These include the amplitude of the power spectrum of the seed perturbation, the primordial curvature perturbation $\calR$, and its running, i.e. the spectral index, which are constrained at $1\sigma$ confidence level by~\cite{Planck:2013jfk}
\begin{equation}
\begin{split}
\log\left( 10^{10}\calP_\calR \right) & = 3.089^{+0.024}_{-0.027} \, ,
\\
n_\calR & = 0.9603 \pm 0.0073 \, ,
\end{split}
\end{equation}
respectively. The primordial perturbation is believed to be generated during cosmic inflation~\cite{inflation} before the onset of the standard hot big bang evolution of the universe. Thus precise observations on the power spectrum of $\calR$ have played a central role in constraining viable models of inflation and the underlying physics of the early universe~\cite{Lyth:2009zz}.

There are, however, a number of compelling hints that we may need more parameters to describe the properties of the primordial perturbations. Especially, the recent detection of the $B$-mode polarization of the CMB by the BICEP2 experiment in the range $30 < \ell < 150$ indicates, if solely from the primordial gravitational waves, a large tensor-to-scalar ratio~\cite{Ade:2014xna},
\begin{equation}
r = 0.20^{+0.07}_{-0.05} \, ,
\end{equation}
although it is still debatable to pin down the fraction of dust polarization contribution~\cite{dust} (see also~\cite{Adam:2014bub}). This is inconsistent with the Planck constraint $r < 0.11$~\cite{Planck:2013jfk}, which may imply that the aspect of inflation was different on the scales relevant for Planck and BICEP2 observations respectively~\cite{whipped}. A more conservative, standard (maybe not sufficient alone though~\cite{Giannantonio:2014rva}) way is to introduce the running of the spectral index,
\begin{equation}
\alpha_\calR \equiv \frac{dn_\calR}{d\log{k}} = \frac{d^2\log\calP_\calR}{d\log{k}^2} \, .
\end{equation}
Reconciling Planck and BICEP2 requires a large negative running, $\alpha_\calR = -0.028 \pm 0.009$~\cite{Ade:2014xna}. In the standard single field inflation, however, running is second order in slow-roll [see \eqref{alpha_singlefield}] and is thus further suppressed compared to the spectral index, which is first order in slow-roll. Moreover, this standard result assumes a single degree of freedom for the inflaton field which is not necessarily the case in the context of model building based on high energy theories such as supergravity~\cite{Lyth:1998xn} -- we may integrate out heavy degrees of freedom systematically to reduce to an effective single field description though~\cite{effective,effective2}. Thus we need to extend the standard result for running to more general cases. In this article, we derive more general formula for the running of the scalar spectral index produced during slow-roll inflation driven by multiple scalar fields.

\section{Momentum dependence in the $\delta{N}$ formalism}

We consider a general class of inflation model with an arbitrary number of scalar fields labeled by $a$, $b$, $c$, $\cdots$,
\begin{equation}\label{action}
\calL = -\frac{1}{2}\gamma_{ab}g^{\mu\nu}\partial_\mu\phi^a\partial_\nu\phi^a - V(\phi^a) \, ,
\end{equation}
where $\gamma_{ab}$ is a generic field space metric and $V(\phi^a)$ is the potential. We do not assume any specific form of $\gamma_{ab}$ or $V(\phi^a)$, except that around the time of horizon crossing slow-roll approximation is valid.

We first note that the $\delta{N}$ formalism~\cite{deltaNold,deltaN,Gong:2002cx} is a geometric relation in the configuration space. For a local patch larger than the horizon scale, we choose two fixed initial and final moments, $t_i$ and $t_f$ respectively. Here, $t_i$ is some time during inflation soon after the observationally relevant modes have passed outside the horizon, and $t_f$ is some time after all isocurvature perturbations have decayed and the curvature perturbation becomes constant. Then, by choosing the flat slicing for the hypersurfaces at $t_i$ and the comoving one at $t_f$, the final comoving curvature perturbation $\calR(t_f)$ is equivalent to the difference in the number of $e$-folds $\delta{N}$ between the comoving and flat slicings from $t_i$ to $t_f$. Moreover, under the slow-roll approximation, we can eliminate the dependence on $\dot\phi^a(t_i)$ so that we can relate $\calR(t_f)$ to the field fluctuations on the initial flat slice $Q^a(t_i)$,
\begin{equation}\label{deltaN}
\calR(t_f) = \delta{N} = N_a(t_i)Q^a(t_i) + \frac{1}{2}N_{ab}Q^a(t_i)Q^b(t_i) + \cdots \, ,
\end{equation}
where $N_a = \partial{N}/\partial\phi^a$ and so on and we have expanded up to second order.

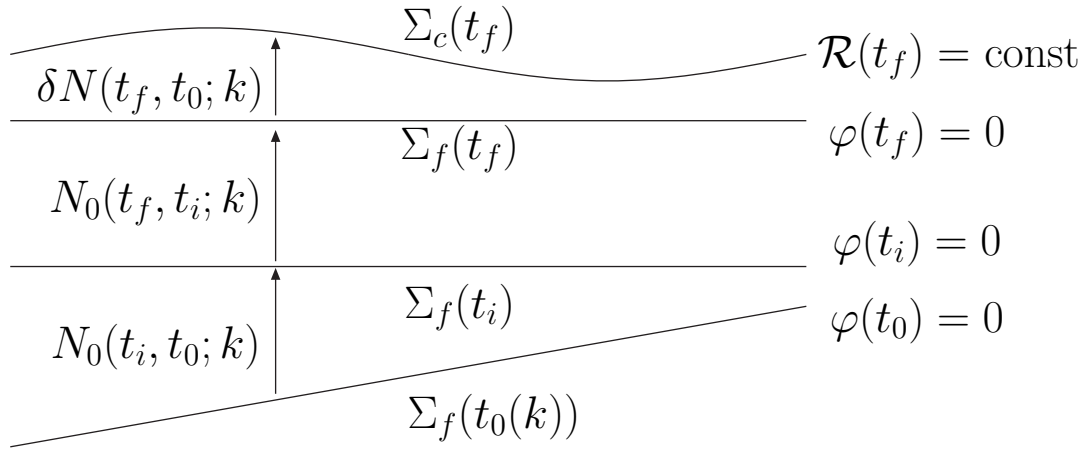
\begin{figure}[t]
\begin{center}
 \begin{picture}(360,160)(0,0)
  \Photon(0,150)(300,150){10}{1}
  \Line(0,125)(300,125)
  \Line(0,70)(300,70)
  \Line(0,2)(300,55)
  \Text(170,160)[]{\Large $\Sigma_c(t_f)$}
   \Text(355,150)[]{\Large $\calR(t_f)=$ const}
  \Text(170,115)[]{\Large $\Sigma_f(t_f)$}
   \Text(343,120)[]{\Large $\varphi(t_f)=0$}
  \Text(170,55)[]{\Large $\Sigma_f(t_i)$}
   \Text(343,79)[]{\Large $\varphi(t_i)=0$}
  \Text(183,13)[]{\Large $\Sigma_f(t_0(k))$}
   \Text(343,50)[]{\Large $\varphi(t_0)=0$}
  \LongArrow(100,127)(100,155)
  \LongArrow(100,72)(100,120)
  \LongArrow(100,22)(100,68)
  \Text(53,137)[]{\Large $\delta N(t_f,t_0;k)$}
  \Text(55,95)[]{\Large $N_0(t_f,t_i;k)$}
  \Text(55,40)[]{\Large $N_0(t_i,t_0;k)$}
 \end{picture}
\end{center}
\caption{A schematic figure showing different moments described in the main text: $t_0(k)$, $t_i$ and $t_f$. We evaluate the final comoving curvature perturbation $\calR(t_f)$ on the hypersurface $\Sigma(t_f)$ with comoving slicing after all isocurvature modes have decayed. We meanwhile choose flat slicing on which the curvature perturbation under this slicing condition, denoted by $\varphi$, of course vanishes. Since $t_i$ and $t_f$ are common to all modes of observational interest, the evolution of $\calR$ is identical between them. But $t_0(k)$ is different for different $k$-modes so the evolution of $\calR$ between $t_0$ and $t_i$ is distinctive for each mode.}
\label{fig:deltaN}
\end{figure}

In this setup, $t_0$ is a fixed, pivot time common to all modes of observational interest and the momentum dependence is not explicit. Previously the $k$-dependence was computed by implicitly assuming that we are considering a function of both $a(t)$ and $k$, with $a$ fixed~\cite{deltaN,Gong:2002cx}. We can, however, make the $k$-dependence more explicit by incorporating the horizon crossing as follows. We consider a certain mode with the momentum $k$, which crosses the horizon at $t_0 < t_i$, i.e. $k = (aH)_0$. Then, until $t_i$ at which we evaluate $Q^a$ for the $\delta{N}$ formalism, each mode gains $k$-dependent $e$-folds as~\cite{Byrnes:2012sc}, from the definition,
\begin{equation}
\Delta{N}_k \equiv \log \left( \frac{a_i}{a_0} \right) \approx \log \left[ \frac{(aH)_i}{k} \right] \, .
\end{equation}
Thus, the field fluctuations at the initial moment for the $\delta{N}$ formalism are related to those at the horizon crossing by Taylor expansion as
\begin{equation}\label{Q_eq1}
Q^a(N_i = N_0 + \Delta{N}_k) = Q^a(N_0) + \Delta{N}_k D_NQ^a(N_0) + \frac{1}{2} \left( \Delta{N}_k \right)^2 D_N^2Q^a(N_0) + \cdots \, ,
\end{equation}
where $N_0$ and $N_i$ are the number of $e$-folds corresponding to $t_0$ and $t_i$, respectively, and $D_N$ is a covariant derivative with respect to $N$. In this manner, the non-trivial $k$-dependence is gained between the {\em tilted} initial moment $t_0(k)$ different for different modes and the conventional initial moment $t_i$ common to all modes. In this sense, we may simply regard $t_i$ as an intermediate reference moment from which we observe identical evolution of each mode until $t_f$: the true initial moment where we can extract $k$-dependence of the curvature perturbation is $t_i(k)$. We can easily understand this point by considering single field inflation, where $\calR$ is conserved between $t_i$ and $t_f$, i.e. $\calR(t_i)=\calR(t_f)$~\cite{conservation}, and non-trivial $k$-dependence is gained between $t_0(k)$ and $t_i$. This is depicted in Figure~\ref{fig:deltaN}. Note that at non-linear level to keep covariance the distinction between the true field fluctuation $\delta\phi^a = \phi^a(t,\mathbi{x}) - \phi^a_0(t)$ and $Q^a$ is important, with the latter in fact being a vector living in the tangent space stemming from $\phi^a_0$~\cite{Gong:2011uw,Elliston:2012ab}. But at linear order $\delta\phi^a$ and $Q^a$ are equivalent and we need not worry about keeping covariance at non-linear level.

To proceed further from \eqref{Q_eq1}, we use the equation of motion for $Q^a$ on very large scales where the space-time metric is that of unperturbed Friedmann-Robertson-Walker one~\cite{Elliston:2012ab},
\begin{align}
D_NQ^a & = w^a{}_bQ^b + \cdots \, ,
\\
w_{ab} & = u_{(a;b)} + \frac{R_{c(ab)d}}{3} \frac{\dot\phi_0^c}{H}\frac{\dot\phi_0^d}{H} \, ,
\\
u_a & = -\frac{V_{;a}}{3H^2} \, ,
\end{align}
where a semicolon denotes a covariant derivative with respect to $\gamma_{ab}$, $R^a{}_{bcd} V^b \equiv V^a{}_{;cd} - V^a{}_{;dc}$ is the Riemann curvature tensor of the field space, and the indices in the parentheses are symmetrized. Then, we can write \eqref{Q_eq1} as
\begin{equation}\label{Q_eq2}
Q^a(N_i) = Q^a + \Delta{N}_k w^a{}_bQ^b + \frac{1}{2} \left( \Delta{N}_k \right)^2 \left[ \left( D_Nw^a{}_b \right)Q^b + w^a{}_bw^b{}_cQ^c \right] + \cdots \, .
\end{equation}
where all terms on the right hand side are evaluated at $N_0$, and
\begin{equation}
D_Nw^{ab} = w^{ab}{}_{;c}\frac{\dot\phi_0^c}{H} \, .
\end{equation}
In this manner, the $k$-dependence is manifest in two ways: through $\Delta{N}_k$ and through $t_0 = t_0(k)$. But as long as we are interested in the leading order expressions, the momentum dependence through the latter does not show up except the one in $H(t_0)$.

\section{Running of the spectral index}

Now we can readily compute the running of the spectral index. Moving to the Fourier space, \eqref{deltaN} becomes
\begin{equation}
\calR_k(t_f) = N_aQ_k^a(t_i) + \cdots \, .
\end{equation}
Then the power spectrum $\calP_\calR$ is given by
\begin{equation}\label{PR}
\left\langle \calR_k(t_f)\calR_q(t_f) \right\rangle = (2\pi)^3 \delta^{(3)}(\mathbi{k}+\mathbi{q}) \frac{2\pi^2}{k^3}\calP_\calR(k) = N_a(t_i)N_b(t_i) \left\langle Q_k^a(t_i)Q_q^b(t_i) \right\rangle \, .
\end{equation}
Using \eqref{Q_eq2}, we can write the two-point correlation function of $Q^a$ at $t_i$ in terms of that at the time of horizon crossing $t_0$ as
\begin{align}\label{PQ}
\left\langle Q_k^a(t_i)Q_q^b(t_i) \right\rangle & = \left\langle Q_k^a(t_0)Q_q^b(t_0) \right\rangle + 2\Delta{N}_kw^a{}_c \left\langle Q_k^b(t_0)Q_q^c(t_0) \right\rangle 
\nonumber\\
& \quad + \left( \Delta{N}_k \right)^2 \left[ \left( D_Nw^a{}_c \right) \left\langle Q_k^b(t_0)Q_q^c(t_0) \right\rangle + w^a{}_cw^c{}_d \left\langle Q_k^b(t_0)Q_q^d(t_0) \right\rangle + w^a{}_cw^b{}_d \left\langle Q_k^b(t_0)Q_q^d(t_0) \right\rangle \right] \, ,
\end{align}
where close to horizon crossing 
\begin{equation}
\left\langle Q_k^aQ_q^b \right\rangle = (2\pi)^3\delta^{(3)}(\mathbi{k}+\mathbi{q}) \frac{H^2}{2k^3} \left( \gamma^{ab} + \varepsilon^{ab} \right) \, ,
\end{equation}
so that to leading order the standard power spectrum for a very light scalar field is reproduced. Here, $\varepsilon^{ab}$ is to leading order $\calO(\epsilon)$ and is slowly varying~\cite{deltaN}. Thus its derivatives with respect to $\log{k}$ always give rise to sub-leading contributions. But the derivatives on $H^2(t_0)$ are of leading order. The rest leading order contributions are entirely from those multiplied by $\Delta{N}_k$.

From \eqref{PR} and \eqref{PQ}, we can easily compute $\calP_\calR$ to leading order
\begin{equation}
\calP_\calR(k) = \frac{N_a(t_i)N_b(t_i)}{(2\pi)^2} H^2 \gamma^{ab} \, ,
\end{equation}
and we can then straightly obtain the spectral index as
\begin{equation}\label{index}
n_\calR-1 \equiv \frac{D\log\calP_\calR}{d\log{k}} = -2\epsilon - 2 \frac{N_aN_bw^{ab}}{N_cN^c} \, ,
\end{equation}
which are in agreement with the known result. Now we can proceed further to find the running of the spectral index as
\begin{equation}\label{running}
\alpha_\calR \equiv \frac{Dn_\calR}{d\log{k}} = 4\epsilon^2 - 2\epsilon\eta + \frac{N_aN_b}{N_cN^c} \left( 8\epsilon w^{ab} + 4w^a{}_cw^{bc} - 2D_Nw^{ab} \right) - \left( n_\calR-1 \right)^2 \, ,
\end{equation}
where $\eta \equiv \dot\epsilon/(H\epsilon)$. Note that $\alpha_\calR$ in this generic context with \eqref{action} was obtained in~\cite{Gong:2002cx,others} in a different form. We can of course check the results are consistent, e.g. for~\cite{Gong:2002cx} by using (47) - (56) there. While one can extract valuable information on, for example, geometry via projections of the potential derivatives onto certain directions~\cite{Gong:2002cx}, the merit of \eqref{running} is twofold on top of computational convenience. First, in deriving \eqref{running} we could clearly see where the scale dependence is contained. Previously this point was somewhat obscure. Second, we can readily distinguish the origin of each contribution in \eqref{running}: as emphasized before, the first two terms are coming from the derivatives of $H^2(t_0)$ in the power spectrum, thus from imperfect de Sitter expansion of the background. Meanwhile the rest contributions are the terms multiplied by $\Delta{N}_k$, and thus are originated from the super-horizon evolution of the field fluctuations from the horizon crossing until the initial moment for the $\delta{N}$ formalism.

As a check of our result \eqref{running}, we consider the well known result in single field inflation. In this case, we can reduce
\begin{align}
w_{ab} & \to 2\epsilon_V - \eta_V \, ,
\\
D_Nw^{ab} & \to \xi_V^2 + 8\epsilon_V^2 - 6\epsilon_V\eta_V \, ,
\end{align}
where the slow-roll parameters in terms of the potential and its derivatives are defined by
\begin{equation}
\begin{split}
\epsilon_V & \equiv \frac{\mpl^2}{2} \left( \frac{V'}{V} \right)^2 \, ,
\\
\eta_V & \equiv \mpl^2\frac{V''}{V} \, ,
\\
\xi_V^2 & \equiv \mpl^4 \frac{V'V'''}{V^2} \, .
\end{split}
\end{equation}
We can also write $\eta$ in terms of these potential slow-roll parameters as $\eta \approx 4\epsilon_V-2\eta_V$. Then, from \eqref{index} and \eqref{running} we can find
\begin{equation}\label{alpha_singlefield}
\alpha_\calR = -2\xi_V^2 - 24\epsilon_V^2 + 16\epsilon_V\eta_V \, .
\end{equation}
This is the standard result for the running in single field inflation.

\section{Conclusions}

In this article, we have considered a generic class of multi-field slow-roll inflation that can accommodate large numbers of multi-field inflation models, and have calculated the running of the spectral index. The result \eqref{running} is a combination of standard slow-roll parameters as well as geometric terms that include the field space curvature tensor and its derivatives. One possible source of a large value of running is the field space curvature. For example, consider $R_{abcd} \sim e^{b(\phi)}$ which may arise in e.g. the Einstein frame action of generic scalar-tensor theories including variants of multi-field inflation models with non-minimal coupling~\cite{multi-nonminimal}, where the kinetic sector of a two-field action becomes
\begin{equation}
\calL \supset -\frac{1}{2}(\partial_\mu\phi)^2 - \frac{1}{2}e^{2b(\phi)}(\partial_\mu\chi)^2 \, .
\end{equation}
This term is, however, multiplied by essentially $\mpl^2(V'/V)^2\ll1$ so may be difficult to observe. Constructing a model of inflation, driven by no matter single or multi-field, that can result in large running seems to remain an open and challenging question.

\subsection*{Acknowledgements}

I thank Eiichiro Komatsu and Hyun Min Lee for helpful conversations. 
I am also grateful to the Munich Institute for Astro- and Particle Physics of the DFG cluster of excellence ``Origin and Structure of the Universe'' for hospitality while this work was finished.
I acknowledge the Max-Planck-Gesellschaft, the Korea Ministry of Education, Science and Technology, Gyeongsangbuk-Do and Pohang City for the support of the Independent Junior Research Group at the Asia Pacific Center for Theoretical Physics. I am also supported by a Starting Grant through the Basic Science Research Program of the National Research Foundation of Korea (2013R1A1A1006701).


\begin{thebibliography}{}


\bibitem{Ade:2013zuv} 
  P.~A.~R.~Ade {\it et al.}  [Planck Collaboration],
  Astron.\ Astrophys.\  {\bf 571}, A16 (2014)
  [arXiv:1303.5076 [astro-ph.CO]].


\bibitem{Planck:2013jfk} 
  P.~A.~R.~Ade {\it et al.}  [Planck Collaboration],
  Astron.\ Astrophys.\  {\bf 571}, A22 (2014)
  [arXiv:1303.5082 [astro-ph.CO]].
    

\bibitem{inflation}
  A.~H.~Guth,
  Phys.\ Rev.\  D {\bf 23}, 347 (1981)~;
  A.~D.~Linde,
  Phys.\ Lett.\  B {\bf 108}, 389 (1982)~;
  A.~Albrecht and P.~J.~Steinhardt,
  Phys.\ Rev.\ Lett.\  {\bf 48}, 1220 (1982).


\bibitem{Lyth:2009zz} 
See e.g. 
  D.~H.~Lyth and A.~R.~Liddle,
  ``The primordial density perturbation: Cosmology, inflation and the origin of structure,''
  Cambridge, UK: Cambridge Univ. Pr. (2009) 497 p


\bibitem{Ade:2014xna} 
  P.~A.~R.~Ade {\it et al.}  [BICEP2 Collaboration],
  Phys.\ Rev.\ Lett.\  {\bf 112}, 241101 (2014)
  [arXiv:1403.3985 [astro-ph.CO]].


\bibitem{dust}
  M.~J.~Mortonson and U.~Seljak,
  JCAP {\bf 1410}, no. 10, 035 (2014)
  [arXiv:1405.5857 [astro-ph.CO]]~;
  R.~Flauger, J.~C.~Hill and D.~N.~Spergel,
  JCAP {\bf 1408}, 039 (2014)
  [arXiv:1405.7351 [astro-ph.CO]]~;
  M.~Cortes, A.~R.~Liddle and D.~Parkinson,
  arXiv:1409.6530 [astro-ph.CO].  


\bibitem{Adam:2014bub} 
  R.~Adam {\it et al.}  [Planck Collaboration],
  arXiv:1409.5738 [astro-ph.CO].


\bibitem{whipped}
  D.~K.~Hazra, A.~Shafieloo, G.~F.~Smoot and A.~A.~Starobinsky,
  JCAP {\bf 1406}, 061 (2014)
  [arXiv:1403.7786 [astro-ph.CO]]~;
  D.~K.~Hazra, A.~Shafieloo, G.~F.~Smoot and A.~A.~Starobinsky,
  Phys.\ Rev.\ Lett.\  {\bf 113}, 071301 (2014)
  [arXiv:1404.0360 [astro-ph.CO]].
  

\bibitem{Giannantonio:2014rva} 
  T.~Giannantonio and E.~Komatsu,
  Phys.\ Rev.\ D {\bf 91}, no. 2, 023506 (2015)
  [arXiv:1407.4291 [astro-ph.CO]].


\bibitem{Lyth:1998xn} 
See e.g. 
  D.~H.~Lyth and A.~Riotto,
  Phys.\ Rept.\  {\bf 314}, 1 (1999)
  [hep-ph/9807278].


\bibitem{effective}
  A.~J.~Tolley and M.~Wyman,
  Phys.\ Rev.\ D {\bf 81}, 043502 (2010)
  [arXiv:0910.1853 [hep-th]]~;
  A.~Achucarro, J.~-O.~Gong, S.~Hardeman, G.~A.~Palma and S.~P.~Patil,
  Phys.\ Rev.\ D {\bf 84}, 043502 (2011)
  [arXiv:1005.3848 [hep-th]]~;
  A.~Achucarro, J.~-O.~Gong, S.~Hardeman, G.~A.~Palma and S.~P.~Patil,
  JCAP {\bf 1101}, 030 (2011)
  [arXiv:1010.3693 [hep-ph]]~;
  A.~Achucarro, J.~-O.~Gong, S.~Hardeman, G.~A.~Palma and S.~P.~Patil,
  JHEP {\bf 1205}, 066 (2012)
  [arXiv:1201.6342 [hep-th]].


\bibitem{effective2}
See also e.g.
  S.~Cremonini, Z.~Lalak and K.~Turzynski,
  JCAP {\bf 1103}, 016 (2011)
  [arXiv:1010.3021 [hep-th]]~;
  A.~Avgoustidis, S.~Cremonini, A.~C.~Davis, R.~H.~Ribeiro, K.~Turzynski and S.~Watson,
  JCAP {\bf 1206}, 025 (2012)
  [arXiv:1203.0016 [hep-th]].


\bibitem{deltaNold}
  A.~A.~Starobinsky,
  JETP Lett.\  {\bf 42}, 152 (1985)
  [Pisma Zh.\ Eksp.\ Teor.\ Fiz.\  {\bf 42}, 124 (1985)]~;
  D.~S.~Salopek and J.~R.~Bond,
  Phys.\ Rev.\ D {\bf 42}, 3936 (1990).
  

\bibitem{deltaN}  
  M.~Sasaki and E.~D.~Stewart,
  Prog.\ Theor.\ Phys.\  {\bf 95}, 71 (1996)
  [astro-ph/9507001]~;
  T.~T.~Nakamura and E.~D.~Stewart,
  Phys.\ Lett.\ B {\bf 381}, 413 (1996)
  [astro-ph/9604103].
  
  
\bibitem{Gong:2002cx} 
  J.~-O.~Gong and E.~D.~Stewart,
  Phys.\ Lett.\ B {\bf 538}, 213 (2002)
  [astro-ph/0202098].
  

\bibitem{Byrnes:2012sc} 
  C.~T.~Byrnes and J.~O.~Gong,
  Phys.\ Lett.\ B {\bf 718}, 718 (2013)
  [arXiv:1210.1851 [astro-ph.CO]].


\bibitem{conservation}
  A.~Naruko and M.~Sasaki,
  Class.\ Quant.\ Grav.\  {\bf 28}, 072001 (2011)
  [arXiv:1101.3180 [astro-ph.CO]]~;
  J.~O.~Gong, J.~c.~Hwang, W.~I.~Park, M.~Sasaki and Y.~S.~Song,
  JCAP {\bf 1109}, 023 (2011)
  [arXiv:1107.1840 [gr-qc]].
  

\bibitem{Gong:2011uw} 
  J.~O.~Gong and T.~Tanaka,
  JCAP {\bf 1103}, 015 (2011)
  [Erratum-ibid.\  {\bf 1202}, E01 (2012)]
  [arXiv:1101.4809 [astro-ph.CO]].


\bibitem{Elliston:2012ab} 
  J.~Elliston, D.~Seery and R.~Tavakol,
  JCAP {\bf 1211}, 060 (2012)
  [arXiv:1208.6011 [astro-ph.CO]].


\bibitem{others}
  E.~D.~Stewart,
  Phys.\ Rev.\ D {\bf 65}, 103508 (2002)
  [astro-ph/0110322]~;
  X.~Gao, T.~Li and P.~Shukla,
  JCAP {\bf 1410}, no. 10, 008 (2014)
  [arXiv:1403.0654 [hep-th]].


\bibitem{multi-nonminimal}
  J.~O.~Gong and H.~M.~Lee,
  JCAP {\bf 1111}, 040 (2011)
  [arXiv:1105.0073 [hep-ph]]~;
  J.~O.~Gong, H.~M.~Lee and S.~K.~Kang,
  JHEP {\bf 1204}, 128 (2012)
  [arXiv:1202.0288 [hep-ph]].


\end{thebibliography}
\end{document}